# Parametrization for Cherenkov light lateral distribution function in Extensive Air Showers

A.A. Al-Rubaiee, O.A. Gress, A.A. Kochanov, K.S. Lokhtin, Y.V. Parfenov and S.I Sinegovsky

*Irkutsk State University, Irkutsk, Russia*
Presenter: V.V. Prosin (kochanov@api.isu.ru), rus-kochanov-A-abs3-he12-poster

The simulation of the Cherenkov light lateral distribution function (LDF) in Extensive Air Showers (EAS) was perfomed with the CORSIKA code in the energy range $10^{13} - 10^{16}$ eV for configuration of the Tunka-25 EAS array. On the basis of this simulation we obtained sets of approximating functions for primary protons, iron nuclei and $\gamma$-quanta for zenith angles $\theta \leq 20°$. The comparison of the calculated Cherenkov light LDF with that measured with the Tunka-25 array has shown an opportunity of primary particle identification and definition of its energy around the "knee".

## 1. Introduction

The studies of the energy spectrum and the chemical composition of primary cosmic rays (CR) in the energy range from $10^{15}$ to $10^{16}$ eV is of a special interest in connection with observed change of the index of primary CR spectrum close to $E \sim 3 \cdot 10^{15}$ eV ("knee"). A method of CR research based on the detection of Cherenkov light from secondary particles, which are produced in EAS cascade processes, intensively developed in last years [1]. The Cherenkov light LDF depends on energy and type of the primary particle, an observation level, a height of the first interaction, a direction of a shower axis and other parameters.

In this work the simulation of Cherenkov light LDF for Tunka-25 array [2, 3] is performed with the CORSIKA 5.61 code [4, 5] using QGSJET [6] and GHEISHA [7] packages for the simulation of hadronic processes and EGS4 code for the simulation of the EAS electromagnetic component and Cherenkov light radiation. We use the following parameters: the wavelength range of Cherenkov radiation $(350 - 600$ nm), threshold energies $(0.3, 0.3, 0.03, 0.03$ GeV), for hadrons, electrons, muons and photons respectively. The calculation has been performed for primary protons, iron nuclei and $\gamma$-quanta in the energy range $10^{13} - 10^{16}$ eV for zenith angles $\theta \leq 20°$. In order to simulate the $10^{17}$ eV shower with CORSIKA code, one needs more than 50 processor's hours of a few GHz facility. This fact enforced us to search a technique to reduce the computation time. In Refs. [8, 9] a function which depends on the distance $R$ from the shower axis and primary energy $E$ was proposed to approximate the simulated Cherenkov light LDF. In present work we used this form for parametrization of the results obtained by CORSIKA EAS simulation and to describe the Cherenkov light EAS measured with the TUNKA-25 array.

## 2. Parametrization of the Cherenkov light LDF

The total number of Cherenkov photons radiated by electrons in EAS is directly proportional to the primary energy $E$ [1]:

$$N_\gamma(E) \approx 3.7 \cdot 10^3 \frac{E}{\beta_t} \approx 4.5 \cdot 10^{10} \frac{E}{10^{15} eV} \qquad (1)$$

where $\beta_t$ is the critical energy that is determined as an energy equal to ionization loss of a particle at the $t$-unit: $\beta_t = \beta_{ion} t_0$. For electron, $\beta_{ion} = 2.2$ MeV $\cdot$ (g $\cdot$ cm$^{-2}$)$^{-1}$, $t_0 = 37$ g$\cdot$ cm$^{-2}$ and $\beta_t = 81.4$ MeV. The experimental measurement of this magnitude is rather difficult, therefore one can use the density of Cherenkov radiation (LDF)- the number of photons per unit of a detector area, which appears as a function of an energy and distance from the shower axis:

$$Q(E,R) = \frac{\Delta N_\gamma(E,R)}{\Delta S}. \quad (2)$$

Direct measurements of Cherenkov light [1] showed that the fluctuation of LDF in EAS is essentially less than that of the total number of photons $N_\gamma$. For parametrization of simulated Cherenkov light LDF, we used the proposed function as a function of the distance $R$ from the shower axis and the energy $E$ of the initial primary particle, which depends on four parameters $a, b, \sigma, r_0$:

$$Q(E,R) = \frac{C\sigma \exp[a - (R/b + (R-r_0)/b + (R/b)^2 + (R-r_0)^2/b^2)]}{b[(R/b)^2 + (R-r_0)^2/b^2 + R\sigma^2/b]}, m^{-2} \quad (3)$$

where $C = 10^3$ m$^{-1}$; $R$ is the distance from the shower axis; $a, b, \sigma, r_0$ are parameters of Cherenkov light LDF. The calculations for LDF are performed in the energy range $10^{13} - 10^{16}$ eV and zenith angles $\theta \leq 20°$. Unlike Ref. [8] we found an energy dependence of the parameters $a, b, \sigma, r_0$ that allows us to calculate the Cherenkov light LDF for any primary energy and fit the LDF which was simulated by CORSIKA code. This energy dependence of LDF parameters is approximated as:

$$k(E) = c_0 + c_1 \lg(E/1eV) + c_2 \lg^2(E/1eV) + c_3 \lg^3(E/1eV). \quad (4)$$

where $k(E) = a, \lg(b/1\text{km}), \lg\sigma, \lg(r_0/(1\text{km}))$, and $c_0, c_1, c_2, c_3$ are coefficients (see Tabl. 1) depending on the type of the primary particles ($p, Fe, \gamma$-quanta) and the zenith angle. In Figure. 1 one can see the results of the simulated Cherenkov light LDF for vertical showers (solid line) and LDF calculated with Eq. (3) (dashed) for (a) $\gamma$-quanta, primary protons and iron nuclei at the energy $10^{14}$ eV; and (b) $\gamma$-quanta in the energy range $(10^{12}-10^{14})$ eV. The characteristics of $\gamma$-quanta in comparison with other primaries like protons and iron nuclei are explained in Figure. 1a. The accuracy of the Cherenkov light LDF approximation for vertical showers for primary protons and $\gamma$- quanta is better than 25 % and close to 20 % for iron nuclei at the distances 80-120 m from the shower axis. For the other distances the accuracy is not less than 10%. And for the interval $R$ of $120 - 300$ m one can find the best agreement of the fitted LDF with that simulated with CORSIKA code.

**Table 1.** Coefficients $c_i$ which determine the energy dependence (Eq. 4) of the parameters $a, b, \sigma, r_0$ for vertical showers.

| $k$ | $c_0$ | $c_1$ | $c_2$ | $c_3$ |
|---|---|---|---|---|
| | | $p$ | | |
| $a$ | $2.222 \cdot 10^1$ | $-9.788 \cdot 10^0$ | $9.465 \cdot 10^{-1}$ | $-2.368 \cdot 10^{-2}$ |
| $b$ | $-5.262 \cdot 10^1$ | $1.113 \cdot 10^1$ | $-7.740 \cdot 10^{-1}$ | $1.784 \cdot 10^{-2}$ |
| $\sigma$ | $-4.971 \cdot 10^{-1}$ | $4.669 \cdot 10^{-1}$ | $-6.560 \cdot 10^{-2}$ | $2.330 \cdot 10^{-3}$ |
| $r_0$ | $-5.866 \cdot 10^1$ | $1.243 \cdot 10^1$ | $-8.761 \cdot 10^{-1}$ | $2.028 \cdot 10^{-2}$ |
| | | $Fe$ | | |
| $a$ | $-4.754 \cdot 10^2$ | $9.184 \cdot 10^1$ | $-5.965 \cdot 10^0$ | $1.328 \cdot 10^{-1}$ |
| $b$ | $8.756 \cdot 10^{-1}$ | $-1.312 \cdot 10^{-1}$ | $9.670 \cdot 10^{-3}$ | $-2.294 \cdot 10^{-4}$ |
| $\sigma$ | $1.487 \cdot 10^1$ | $-3.389 \cdot 10^0$ | $2.487 \cdot 10^{-1}$ | $-6.000 \cdot 10^{-3}$ |
| $r_0$ | $2.286 \cdot 10^1$ | $-4.619 \cdot 10^0$ | $3.097 \cdot 10^{-1}$ | $-7.130 \cdot 10^{-3}$ |
| | | $\gamma$ | | |
| $a$ | $3.348 \cdot 10^2$ | $-8.168 \cdot 10^1$ | $6.499 \cdot 10^0$ | $-0.167 \cdot 10^0$ |
| $b$ | $-0.636 \cdot 10^0$ | $0.245 \cdot 10^0$ | $-0.205 \cdot 10^{-1}$ | $5.549 \cdot 10^{-4}$ |
| $\sigma$ | $-0.650 \cdot 10^0$ | $0.761 \cdot 10^{-1}$ | $-0.593 \cdot 10^{-2}$ | $1.569 \cdot 10^{-4}$ |
| $r_0$ | $6.059 \cdot 10^1$ | $-1.482 \cdot 10^1$ | $1.203 \cdot 10^0$ | $-0.327 \cdot 10^{-1}$ |

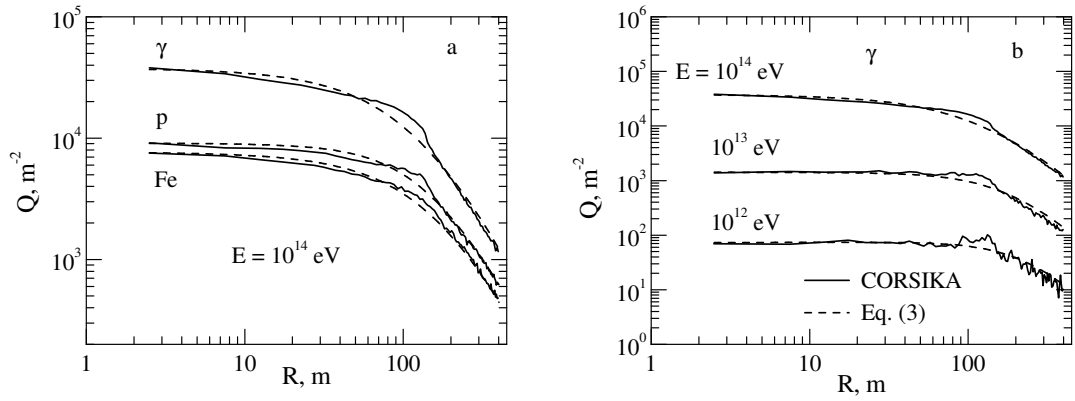

**Figure 1.** Lateral distribution of Cherenkov light which simulated with CORSIKA code (solid lines) and one calculated (Eq. (3)) (dashed lines) for vertical showers initiated (a) – by primary $\gamma$-quanta, protons and iron nuclei at $E = 10^{14}$ eV; (b) – by $\gamma$-quanta in the energy range $10^{12} - 10^{14}$ eV.

## 3. Comparison of the calculations with the Tunka-25 measurements

The wide-angle Tunka-25 Cherenkov array [2, 3] designed for studying the CR spectrum and the mass composition near the "knee" consists of 25 detectors arranged on the square of 340x340 m$^2$ at 675 m above sea level with the distance between detectors of 85 m. The main parameters of EAS measurements are zenith and azimuth angles, shower core location, individual LDF slope parameter $R_0$ and the density of Cherenkov radiation $Q(R)$. The measured Cherenkov light LDF with the Tunka-25 is most sensitive to EAS at the distances $20 - 150$ m from the shower axis depending on the EAS longitudinal development $R_0$.

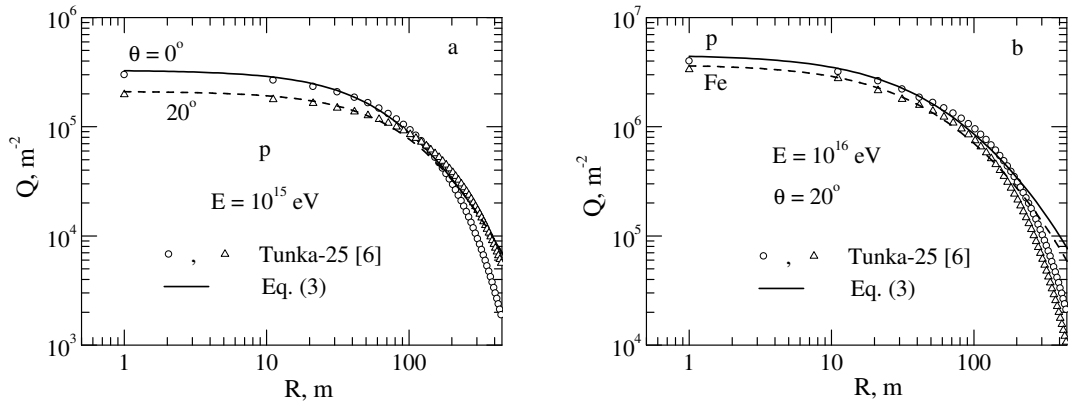

**Figure 2.** Comparison of the calculated Cherenkov light LDF (Eq. (3)) with the data obtained by the Tunka-25 (symbols). (a) – proton, $\theta = 0°$ (solid line), $\theta = 20°$ (dashed); (b) – proton (solid line), iron nuclei (dashed) for $\theta = 20°$ and $E = 10^{16}$ eV.

In Figure. 2a the comparison between the Cherenkov light LDF which was calculated with Eq. (3) with that measured with Tunka-25 array is presented. Figure. 2b displays the difference between the primary proton (solid line) and the iron nuclei initiated EAS (dashed) in comparison with the Tunka-25 at the energy $10^{16}$ eV for $\theta = 20°$. One may see reasonable agreement between the calculated LDF and measurements of LDF at the distances $10 - 200$ m from the shower axis. These figures demonstrate the possibility for reconstruction the type of the EAS primary particles. The calculated Cherenkov light LDF in Figure. 2 for vertical showers slightly differs from the LDF measured with the Tunka-25: at the distance interval 10-200 m the distinction is about 15% for primary proton at the energy $E = 10^{15}$ eV, for $\theta = 20°$ the accuracy is about 10-20%. For the iron initiated showers the accuracy of the calculated Cherenkov light LDF is close to the accuracy for primary protons.

## 4. Conclusion

In this work the calculations of the lateral distribution function of Cherenkov light in EAS initiated by primary protons, iron nuclei and $\gamma$- quanta, were performed in the energy interval $10^{13} - 10^{16}$ eV. The CORSIKA simulation of the Cherenkov light LDF in EAS is performed for configuration of the Tunka-25 EAS array. Using results of this simulation we obtained the parameters of LDF as a functions of the primary energy for different primary particles and zenith angles. The comparison of the calculated Cherenkov light LDF with that measured with the Tunka-25 array demonstrates the possibility to identify the primary particles and to determine their energies around the knee. The main advantage of the given approach consists of the possibility to make a library of LDF samples which could be utilized for analysis of real events which detected with the EAS array and reconstruction of the primary cosmic rays energy spectrum and mass composition.

## References

[1] G.B. Khristiansen, G.V. Kulikov, Y.A. Fomin, Kosmicheskoe izluchenie sverkhvysokoi energii. M. 1975. 256 p.
[2] N. Budnev, D. Chernov, V. Galkin et al., Proc. 27 ICRC, Hamburg, 7-15 Aug 2001. P. 581-584.
[3] N.M. Budnev, R.V. Vasilev, R. Wishnevski, et al., Proc. 28 RCRC, Moscow, 7-11 Jun 2004. P. 1206-1209.
[4] J. Knapp, D. Heck, S.J. Sciutto, et al. // Astropart. Phys. 2003. V. 19. P. 77-99.
[5] D. Heck, J. Knapp, J.N. Capdevielle, et al. // CORSIKA: A Monte Carlo Code to Simulate Extensive Air Showers. Report FZKA 6019. Forschungszentrum Karlsruhe. 1998. 90 p.
[6] S. Ostapchenko, hep-ph/0412332, 2004 ; astro-ph/0412591, 2004.
[7] D. Heck, R. Engel, Proc. 28th ICRC, Tsukuba, Japan, 2003. P. 279-282.
[8] L. Alexandrov, S.Cht. Mavrodiev, A. Mishev , Proc. 27 ICRC, Hamburg, 7-15 Aug 2001. P. 257-260.
[9] S. Mavrodiev, A. Mishev, J. Stamenov, Proc. 28th ICRC, Tsukuba, Japan, 2003. P. 163-174.